\documentclass[12pt]{article}

\usepackage[body={17.5cm, 21cm},right=2cm]{geometry}

\usepackage[utf8]{inputenc}
\usepackage{hyperref}

\usepackage{color}
\usepackage{graphicx}
\usepackage{epsf}
\usepackage{graphicx,epsfig}
\usepackage{amsmath}
\usepackage{amssymb}
\usepackage{epsfig}
\usepackage{cite}
\usepackage{color,colordvi}
\newcommand{\be}{\begin{eqnarray}}
\newcommand{\ee}{\end{eqnarray}}
\newcommand{\bi}{\begin{itemize}}
\newcommand{\ei}{\end{itemize}}

\newcommand{\bx}{{\vec{x}}}
\newcommand{\del}{\partial}

\let\latexcirc=\circ
\newcommand{\ccirc}{\mathbin{\mathchoice
  {\xcirc\scriptstyle}
  {\xcirc\scriptstyle}
  {\xcirc\scriptscriptstyle}
  {\xcirc\scriptscriptstyle}
}}
\newcommand{\xcirc}[1]{\vcenter{\hbox{$#1\latexcirc$}}}
\let\circ\ccirc

\newcounter{hran}

\def\MSbar{\relax\ifmmode\overline{\rm MS}\else{$\overline{\rm MS}${ }}\fi}

\def\d{\rm d}

\def\d{{\rm d}}

\def\vq{\vec{q}}
\def\vx{\vec{x}}
\def\B{\vec{B}}

\def\bx{{\vec{x}}}
\def\bk{{\vec{k}}}
\def\bnabla{{\vec{\nabla}}}
\def\bv{{\vec{v}_{\rm c}}}
\def\t{\tau}

  \def\vn{\vec{n}}

 \def\vx{\vec{ x}} 
\def\vk{\vec{k}}
\def\vq{\vec{q}}
\def\vy{\vec{y}}
\def\vv{\vec{v}_{\rm p}}
\def\dT{\dot{T}}

\def\Tr{{\rm Tr}}

\def\simlt{\stackrel{<}{{}_\sim}}

\numberwithin{equation}{section}
\begin{document}
\thispagestyle{empty}
\vspace{5mm}
\vspace{0.5cm}
\begin{center}

\def\thefootnote{\fnsymbol{footnote}}

{\Large \bf 
Testing the Origin of Cosmological Magnetic Fields \\
through the Large-Scale Structure Consistency Relations\\
\vspace{0.25cm}	
}
\vspace{1.5cm}
{\large  
P. Berger$^{a}$, A. Kehagias$^{a,b}$ and A. Riotto$^{a}$
}
\\[0.5cm]

\vspace{.3cm}
{\normalsize {\it  $^{a}$ Department of Theoretical Physics and Center for Astroparticle Physics (CAP)\\ 24 quai E. Ansermet, CH-1211 Geneva 4, Switzerland}}\\

\vspace{.3cm}
{\normalsize { \it $^{b}$ Physics Division, National Technical University of Athens, \\15780 Zografou Campus, Athens, Greece}}\\

\vspace{.3cm}


\end{center}

\vspace{3cm}

\hrule \vspace{0.3cm}
{\small  \noindent \textbf{Abstract} \\[0.3cm]
\noindent 
We study the symmetries of the post-recombination cosmological magnetohydrodynamical equations which describe the evolution of dark matter,  baryons and magnetic fields  
in a self-consistent way. This is done both at the level of fluid equations and of  Vlasov-Poisson-Maxwell equations in phase space.  We  discuss some consistency relations for  the soft limit of the $(n + 1)$-correlator functions involving magnetic fields and matter  overdensities. In particular, we stress that any violation of such
consistency relations at equal-time would point towards  an inflationary origin of the magnetic field.

\vspace{0.5cm}  \hrule
\vskip 1cm

\def\thefootnote{\arabic{footnote}}
\setcounter{footnote}{0}


\baselineskip= 18pt

\newpage 
\section{Introduction}\pagenumbering{arabic}
Magnetic fields appear to be ubiquitous in astrophysical and cosmological environments \cite{ruth}.  The largest 
observable magnetic fields are found inside the atmospheres of galaxy
clusters with a strength  of the order of the $\mu$G in the core regions. Magnetic fields 
are found   at even larger   scales; they   are not associated to   collapsing or gravitationally bound structures, but are 
 coherent on scales greater than the largest known structures (about $10^2$ Mpc or even the Hubble radius) and  they permeate the whole universe. The magnetic fields at such different scales are possibly produced by amplification of preexisting
magnetic fields via the dynamo mechanism \cite{dynamo} associated to a subsequent compression and turbulent flows generated during the formation of  the large-scale structure. This mechanism needs some primordial seeds which may be either produced during some era much earlier than structure formation or during the formation of the first objects. In the first category falls the so-called inflationary magnetogenesis  \cite{widrow,ratra} where magnetic fields coherent on very large (super-Hubble) cosmological scales  are generated thanks to a  coupling which breaks the 
conformal invariance of electromagnetism \cite{wreview}. 

In order to describe Dark Matter (DM), baryons, and magnetic fields  
in a self-consistent way and within  a cosmological set-up, one needs 
to numerically solve the full set of equations
of cosmological MagnetoHydroDynamics (MHD). 
Given the highly non-linear nature of these equations, one relies more and more on numerical simulations (see for instance Ref. \cite{dk} for a recent work) and it seems clear by now   that a large-scale primordial field
 is needed to explain the presence of the observed
magnetic fields in galaxy clusters  \cite{dolag}. 

In this paper we wish to take a different, albeit modest, path to provide some information about  the interplay between the large-scale structure formation, the magnetic fields, and their origin. Our method is based on symmetries.
There is no doubt that symmetries play a crucial  role in high energy physics allowing, for instance,  to
derive non-perturbative identities among correlation functions which remain valid even after renormalization \cite{WT}. 

Symmetries are also relevant in the cosmological setting. For instance,   during inflation \cite{lrreview}  the de Sitter isometry group acts  as conformal group  on $\mathbb{R}^3$ when the fluctuations are on super-Hubble scales. In such a  regime,  the SO(1,4) isometry
of the de Sitter background is realized as conformal symmetry of the flat $\mathbb{R}^3$ sections  and   correlators are constrained by conformal invariance \cite{antoniadis,creminelli1,us1,us2,us3}. This  applies in the case in which the cosmological perturbations are generated 
by light scalar fields other than the inflaton (the field that drives inflation). In the opposite case in which the inflationary perturbations
originate from  only one degree of freedom,  one can construct  consistency relations  relating
an $n$-point function to an $(n+1)$-point function where the additional leg is a soft curvature
perturbation, the Goldstone boson associated with a non-linearly realized symmetry. This is possible because
the symmetries shift the comoving curvature perturbation $\zeta$ in a non-linear way, precisely  like a Goldstone boson\cite{creminelli2,hui,baumann1,baumann2}. 
 
 The same arguments can be applied to the case of the large-scale structure \cite{noi1,PietroniPeloso1}. Indeed,  
there are large-scale structure observables, such as the DM peculiar
velocity, which are shifted in a non-linear way under some symmetry transformation, while
the same transformation shifts the density contrasts  only linearly. This gives rise to new kinds of consistency
relations  where the Goldstone boson  is the peculiar velocity.
The power of the consistency relations is that they are non-perturbative and this is very useful when  studying the large-scale structure, where one has to deal with  small and  non-linear scales. 
The consistency relations in the large-scale structure have been recently  
 the subject of an intense activity \cite{c1,acc,noi2,c2,PietroniPeloso2,val1,c3,val2,noi3} as they 
 have the virtue of being true also for the galaxy overdensities, independently of the bias
between galaxies and DM. As such, they may serve as a guidance in building up the bias theory, that is  the
theory connecting the observed galaxy correlators to the underlying DM ones. 

In this paper we wish to show that the symmetry arguments can be adopted to learn something about the correlators
among the magnetic fields, the DM and the baryons in the limit in which one of the momenta is soft. Apart from this restriction,  all the other momenta can correspond to non-perturbative non-linear scales. We will see that there is a nice by-product of this sort of soft-pion theorem applied to magnetic fields: the consistency relations are violated at equal-time in the case in which the seeds of the primordial magnetic fields are of  inflationary origin. Testing the violation of the consistency relation therefore provides a way to test the ultimate origin of the
cosmological magnetic fields we observe in the universe.

The paper is organized as follows. In section 2 we discuss the symmetries of cosmological MHD  in the fluid approximation, while the full set of symmetries  
in phase space is described in the Appendix A. 
In section 3 we derive the consistency relations involving the magnetic fields, and in section 4 we discuss the violation of the consistency relations and its implications. Finally, we conclude in section 5.

\section{Symmetries of cosmological magnetohydrodynamics}
Our starting point is a system  containing the same physical content of the N-body simulations: DM particles treated as a fluid,  a baryon component that  behaves like an ideal  conducting plasma, and the magnetic field in 
the post-recombination era. Moreover, our considerations are valid even in the presence of a cosmological constant, that is in a $\Lambda$CDM universe.
Of course, the fluid approximation breaks down at sufficiently small scales where, for instance, the single stream approximation is no longer valid. However, the symmetries we are going to employ extend also to  the Vlasov-Poisson-Maxwell equations in phase space as we show in the Appendix A. Our considerations are therefore valid beyond the fluid approximation which we assume in the bulk of the paper for simplicity.

To treat all the components in a self-consistent way, the equations of ideal MHD have to be solved together with the DM fluid equations
in a fully cosmological setting. The baryonic plasma is taken to be collisional, with small resistivity so that the equations MHD can be applied. 
We indicate by 

\be
\delta_{\rm c}(\bx,\t)=\frac{\rho(\bx,\t)}{\overline{\rho}_{\rm c}}-1,\,\,\, \delta_{\rm p}(\bx,\t)=\frac{\rho_{\rm p}(\bx,\t)}{\overline{\rho}_{\rm p}}-1
\ee
the DM and the plasma overdensities  defined  over the mean DM density $\overline{\rho}_{\rm c}$ and the plasma density $\overline{\rho}_{\rm p}$, respectively. In addition, we define by 
$\bv(\bx,\t)$ and $\vv(\bx,\t)$ the  peculiar velocities of the DM  fluid and plasma\footnote{Notice that the plasma velocity should be intended as the mean velocity of ions and electrons, see the Appendix A for more details.}, respectively, and by $\Phi(\bx,\t)$  the gravitational  potential due to density 
fluctuations. 

The equations which specify the dynamics of DM, plasma, gravity and the magnetic field $\B(\vx,\t)$ are

\begin{itemize}
\item the DM mass conservation
\be
\frac{\partial \delta_{\rm c}(\bx,\t)}{\partial \t}+\bnabla(1+\delta_{\rm c}(\bx,\t))\bv(\bx,\t)=0\label{fl1},
\ee
\item the DM momentum conservation
\be
 \frac{\partial \bv(\bx,\t)}{\partial \t}+{\cal{H}}(\t)\bv(\vx,\t)
 +[\bv(\vx,\t)\cdot \bnabla]\bv(\bx,\t)=-\bnabla\Phi(\bx,\t),\label{fl2}
\ee
\item the Poisson equation
\be
\nabla^2\Phi(\bx,\t)=\frac{3}{2} {\cal{H}}^2(\tau)
\Big{(}\Omega_{\rm c}\delta_{\rm c}(\bx,\t)+\Omega_{\rm p}\delta_{\rm p}(\bx,\t)\Big{)}, \label{fl3}
\ee
\item the plasma mass conservation
\be
  \frac{\del \delta_{\rm p}(\bx,\t)}{\del \t}+\bnabla  (1+\delta_{\rm p}(\bx,\t)) \vv(\bx,\t)=0,\label{m1}
\ee
\item the plasma momentum  conservation
\begin{eqnarray}
&&\rho_{\rm p}(\bx,\t)\left[
 \frac{\del \vv (\bx,\t)}{\del \t}+ {\cal{H}}(\t)\vv(\vx,\t)+(\vv(\bx,\t)\cdot\bnabla)\vv(\bx,\t)  +\bnabla \Phi(\bx,\t)\right]\nonumber\\
 &=&
 \frac{1}{\mu}(\vec{B}(\bx,\t)\cdot\bnabla)\vec{B}(\bx,\t)
 -\frac{1}{2\mu}\bnabla B^2(\bx,\t)-\bnabla P_{\rm p}(\bx,\t),\label{m2}
\end{eqnarray}
\item the induction equation
\be
\frac{\del \vec{B}(\bx,\t)}{\del \t}+2{\cal H}\B-\bnabla\times (\vv(\bx,\t)\times \vec{B}(\bx,\t))=0,\label{m3}
 \ee
\item Gauss' law for magnetism
 \be
 \bnabla\cdot\B(\bx,\t)=0 \label{m4}.
 \ee
 \end{itemize}
 We have  denoted by $\bx$  the comoving spatial coordinates, $\d\tau= \d t/a$  the conformal time, $a$ the scale factor
in the FRW metric,  
${\cal{H}}=\d\ln a /\d\t$  the conformal expansion rate, $\vec{B}(\bx,\t)$ the  magnetic field  strength,
and by $P_{\rm p}(\bx,\t)$ the plasma pressure. Finally,  $
\Omega_{\rm c}=8\pi G\overline \rho  a^2/3{\cal{H}}^2$ and $\Omega_{\rm p}=8\pi G\overline \rho _{\rm p} a^2/3{\cal{H}}^2$
are the density parameters, and $\mu$ is the permeability of the plasma.

We wish now to show that the  above equations are invariant  under the following set of  transformations
\allowdisplaybreaks
 \begin{align}
 \t'=\t,&~~~\bx'=\bx+\vec{n}(\t),\label{ag1}\\
 \delta'_{\rm c}(\bx,\t)&=\delta_{\rm c}(\bx',\t'),\label{dg1}\\
 \bv'(\bx,\t)&=\bv(\bx',\t')-\dot{\vec{n}}(\t), \label{vg1}\\
\delta'_{\rm p}(\bx,\t)&=\delta_{\rm p}(\bx',\t'),\label{mt1}\\
\vv'(\bx,\t)&=\vv(\bx',\t')-\dot{\vec{n}}(\t), \label{mt2}\\
\vec{B}'(\vx,t)&=\vec{B}(\vx',\t'),\label{tB}\\
P_{\rm p}'(\vx,t)&=P_{\rm p}(\vx',\t'),\\
 \Phi'(\bx,\t)&=\Phi(\bx',\t')+\left(\ddot{\vec{n}}(\t)+\mathcal{H}(\t) \dot{\vec{n}}(\t)\right)\cdot \bx, \label{gfg1}
 \end{align}
 where $\vn(\t)$ is an arbitrary time-dependent vector. In other words, if  $\delta_{\rm c}(\bx,\t),\bv(\bx,\t),\cdots$ are solutions, then also $\delta'_{\rm c}(\bx,\t),\bv'(\bx,\t),\cdots$ are  solutions of the cosmological MHD equations. 
 The  invariance of the Eqs. (\ref{fl1}-\ref{m4}) under the transformations (\ref{ag1}-\ref{gfg1}) can be easily checked by noticing that temporal and spatial derivatives transform as 
 \be
 \label{dg}
\left.\frac{\partial}{\partial \t}\right|_{\vx}=\left.\frac{\partial}{\partial \t'}\right|_{\vx'}+\dot{\vec{n}}\cdot \bnabla'\, , ~~~\bnabla=\bnabla'.
\ee
This implies that the operators
\be
D_\t^{v_{\rm c}}=\frac{\partial}{\partial \t}+\bv(\bx,\t)\cdot \bnabla, ~~~D_\t^{v_{\rm p}}=\frac{\partial}{\partial \t}+\vv(\bx,\t)\cdot \bnabla
\ee 
are left invariant. Indeed, the invariance of Eqs. (\ref{fl1}-\ref{fl3}) has  been proven in Ref. \cite{noi1}, but we repeat
here for the benefit of the reader. We start with the DM momentum conservation (\ref{fl2})

\begin{align}
\label{fl22}
0&=\frac{\partial \bv'(\bx,\t)}{\partial \t}+{\cal{H}}(\t)\bv'(\bx,\t)
 +[\bv'(\bx,\t)\cdot \bnabla]\bv'(\bx,\t)+\bnabla\Phi'(\bx,\t)\nonumber \\
 &=\frac{\partial \bv(\bx',\t')}{\partial \t'}-\ddot{\vec{n}}+{\cal{H}}(\t')\bv(\bx',\t')
 -{\cal{H}}(\t')\dot{\vec{n}}
 +[\bv(\vx',\t')\cdot \bnabla']\bv(\bx',\t')+\bnabla\Phi'(\bx,\t)\nonumber \\
 &=-\ddot{\vec{n}}-
{\cal{H}}(\t')\dot{\vec{n}}-\bnabla'\Phi(\bx',\t')+\bnabla\Phi'(\bx,\t), 
\end{align}
 from which we deduce the transformation 
\be
\bnabla\Phi'(\bx,\t)=\bnabla'\Phi(\bx',\t')+\ddot{\vec{n}}+
{\cal{H}}(\t')\dot{\vec{n}}
\ee
or
\be
\Phi'(\bx,\t)=\Phi(\bx',\t')+\left(\ddot{\vec{n}}+
{\cal{H}}(\t')\dot{\vec{n}}\right)\cdot \bx.
\ee
The invariance of the Poisson equation (\ref{fl3}) automatically follows. The DM and plasma mass conservations (\ref{fl1}) and (\ref{m1}) respectively also follow from rewriting them  as
\be
D_\t^{v_{\rm c}}\delta(\bx,\t)+(1+\delta(\bx,\t))\bnabla\cdot\bv(\bx,\t)=0,\,\,\,D_\t^{v_{\rm p}}\delta_{\rm p}(\bx,\t)+(1+\delta_{\rm p}(\bx,\t))\bnabla\cdot\vv(\bx,\t)=0.
\ee
The invariance of the plasma momentum conservation (\ref{m2})  is proved similarly to what is done for the DM momentum conservation equation (\ref{fl2}) with the addition that the plasma density contrast $\delta_{\rm p}(\bx,\t)$, the magnetic
field $\B(\bx,\t)$, and the pressure $P_{\rm p}(\bx,\t)$ are scalars under the transformations. The same reasoning applies
to the Gauss' law (\ref{m4}). 

Finally, let us consider the induction equation (\ref{m3}). 
Using the fact that 
\be
\left[\bnabla\times (\vv\times \vec{B})\right]=-(\bnabla\cdot \vv)\B-(\vv\cdot\bnabla) \B+
\vv(\bnabla\cdot \vec{B})+(\vec{B}\cdot\bnabla)\vv
\ee
and exploiting  $\bnabla\cdot \vec{B}=0$, we can recast the induction equation in the form
\begin{align*}
0&=\frac{\del \vec{B}}{\del \t}+2{\cal H}\B
+(\bnabla\cdot \vv)\vec{B}+(\vv\cdot\bnabla) \vec{B}-(\vec{B}\cdot\bnabla)\vv\nonumber\\
&=D_\t^{v_{\rm p}}\vec{B}+2{\cal H}\B+(\bnabla\cdot \vv)\vec{B}-(\vec{B}\cdot\bnabla)\vv,\nonumber
\end{align*}
which is invariant since it involves the invariant derivative $D_{\t}^{v_{\rm p}}$ and gradients of $\vv$.


We close this section with two comments. First, the symmetry of the cosmological MHD equations are correct even though they are highly non-linear. This crucial point will allow us to find  consistency relations involving the magnetic field at any scales. Secondly, if  one introduces back in the plasma momentum conservation equation (\ref{m2}) and the induction equation (\ref{m3}) the terms depending on the shear viscosity and conductivity respectively, our symmetry arguments do not change. This is because the shear viscosity terms are proportional to gradients of the peculiar velocities and the conductivity terms are proportional to the Laplacian of the magnetic field.

\section{Cosmological magnetohydrodynamics consistency relations}
\noindent
In this section we wish to analyze the implications of the symmetry discussed previously.  
Consider the $n$-point correlation function of short modes of the magnetic field. The symmetries of the cosmological MHD equations equations imply, for instance, that 
\be
\label{ward}
\Big<\B'(\,t_1,\vx_1)\cdots \B'(\t_n,\vx_n)\Big>=\Big<\B(\t_1,\vx_1)\cdots \B(\t_n, \vx_n)\Big>=\Big<\B(\tau_1',\vx'_1)\cdots \B(\tau_n',\vx'_n)\Big>.
\ee
The meaning of the Eq. (\ref{ward}) is that
the correlators in the two coordinate systems have to be the same as both $\B(\tau_i,\vx_i)$ and $\B'(\tau_i,\vx_i)=\B(\tau'_i,\vx'_i)$ ($i=1,\cdots,n$) satisfy the equations on motion. 

We can take advantage of this relation in the following way. 
Suppose that the $n$ points are  contained in a sphere of radius  $R$ 
 much smaller than a long wavelength mode of size $\sim 1/q$ and centered at the origin of the coordinates.  Since the cosmological MHD  equations  are invariant under the generic transformation $\t\rightarrow \t$ and  $\bx\rightarrow \bx+{\vec{n}}(\tau)$, and the DM peculiar velocity transforms non-linearly under such a symmetry (like a Goldstone boson)
 \be
 \bv'(\bx,\t)&=\bv(\bx',\t')-\dot{\vec{n}}(\t),
 \ee
this means that we can go to a new coordinate system to cancel (or generate) a long wavelength mode for the DM  velocity perturbation $\vec{v}_{{\rm c}L}(\t,\vec{0})$ just by choosing properly
the vector ${\vec{n}}(\tau)$
\be
{\vec{n}}(\tau)=\int^\tau\d\eta\, \vec{v}_{{\rm c}L}(\eta, \vec{0}) +{\cal O}(qR v_L^2).
\label{long}
\ee
This is true if we may consider a sufficiently long wavelength mode $1/q$ such that
the gravitational potential and its gradient can be considered constant in space on a scale of size $R$. This amounts to removing  the time-dependent but homogeneous gravitational force via a change of
coordinates\footnote{To get convinced about this point, let us suppose that the impact of baryons and the magnetic field onto the
peculiar DM velocity is negligible. If so, one finds that the transformation (in a matter-dominated universe) becomes $ \t'=\t,~~~\bx'=\bx+\int^\tau\d\eta\, \vec{v}_{{\rm c}L}(\eta)=\bx+(1/6)\t^2\bnabla\Phi_L$.
Neglecting the  magnetic field contribution to $\delta_{{\rm c}L}(\vq,\t)$, and therefore to the DM peculiar velocity,  is a good approximation because
 the linear evolution of the DM density contrast $\delta_{{\rm c}L}(\vq,\t)$  receives corrections from   the magnetic field (through the gravitational coupling to the baryon fluid) which are at least  of the order of $(B_{\rm in}/{\rm nG})^2(q\,{\rm Mpc})^2\tau^2$, where $B_{\rm in}$ is the initial condition
for the comoving magnetic field on large scales \cite{al}. For linear scales $1/q\sim 10^{-2}$ Mpc and for $B_{\rm in}\sim$ nG, or less, the corresponding contribution is totally negligible. This is also confirmed by N-body simulation results \cite{dk}.}.

The correlator of the short wavelength modes  of the magnetic field in the background of the long wavelength mode perturbation should satisfy the relation 
\be
\Big<\B(\t_1,\bx_1)\B(\t_2,\bx_2)\cdots\B(\t_n,\bx_n)\Big>_{v_L}=
\Big<\B(\t_1',\bx_1')\B(\t_2',\bx_2')\cdots\B(\t_n',\bx_n')\Big>.
\ee
This equation tells us that  effect of a physical long wavelength DM  velocity perturbation  onto the short modes of the magnetic field 
should be  indistinguishable from the long wavelength mode velocity generated by the transformation with $\bx\rightarrow \bx+{\vec{n}}(\tau)$.
Correlating with the linear long wavelength mode of the DM density contrast in momentum space,  we obtain
\be
\Big< \delta_{{\rm c}L}(\vq,\t)\B({\vk_1},\t_1)\cdots\B(\vk_n,\t_n)\Big>_{q\to 0}=
\Big< \delta_{{\rm c}L}(\vq,\t)
\Big<\B(\vk_1,\t_1)
\cdots\B(\vk_n,\t_n)\Big>_{v_L}\Big>. 
\ee
The variation of the  $n$-point correlator under the infinitesimal transformation  is given 
at first-order in $\delta x^i_a =n^i(\tau_a)$ by
\begin{align}
\delta_n \Big<\B(\t_1,\bx_1)\cdots\B(\t_n,\bx_n)\Big>&\simeq
\int \frac{\d^3\vk_1}{(2\pi)^3}\cdots \frac{\d^3\vk_n}{(2\pi)^3} \Big<\B(\vk_1,\t_1)
\cdots\B(\vk_n,\t_n)\Big>\nonumber \\
&\times \sum_{a=1}^n\delta x^i_a (i k_a^i) e^{i(\vk_1\cdot \vx_1+\cdots\vk_n\cdot \vx_n)}\nonumber \\
&=\int \frac{\d^3\vk_1}{(2\pi)^3}\cdots \frac{\d^3\vk_n}{(2\pi)^3} \Big<\B(\vk_1,\t_1)
\cdots\B(\vk_n,\t_n)\Big>\nonumber \\
&\times \sum_{a=1}^n n^i(\tau_a) (i k_a^i) e^{i(\vk_1\cdot \vx_1+\cdots\vk_n\cdot \vx_n)}.
\end{align}
Therefore we find that 
\begin{eqnarray}
\Big< \delta_{{\rm c}L}(\vq,\t)\B(\vk_1,\t_1)\cdots\B(\vk_n,\t_n)\Big>_{q\to 0}&=&
\Big< \delta_{{\rm c}L}(\vq,\t)\Big<\B(\vk_1,\t_1)\cdots\B(\vk_n,\t_n)\Big>_{v_L}\Big> \nonumber \\
&=&i\sum_{a=1}^n\Big< \delta_{{\rm c}L}(\vq,\t) n^i(\tau_a)\Big>  k_a^i \Big<\B(\vk_1,\t_1)
\cdots\B(\vk_n,\t_n)\Big>.
\end{eqnarray}
In    a $\Lambda$CDM model we have
\begin{align}
\int^\tau\d\eta\, \vec{v}_{{\rm c}L}({\vq},\eta)&=
i\frac{\vq}{q^2}\int^\tau\d\eta\,\frac{\partial}{\partial\eta}\delta_{{\rm c}L}({\vq},\eta)
=i\frac{\vq}{q^2}\delta_{{\rm c}L}(\vq,\tau).
\end{align}
We thus obtain the consistency relation
\be
\fbox{$\displaystyle
\Big< \delta_{{\rm c}L}(\vq,\t)\B(\vk_1,\t_1)\cdots\B(\vk_n,\t_n)\Big>'_{q\to 0}
= -\sum_{a=1}^n  \frac{{\vec q} \cdot \bk_a}{q^2}
\Big<\delta_{{\rm c}L}(q,\tau)\delta_{{\rm c}L}(q,\tau_a)\Big>'
\Big<\B(\vk_1,\t_1)
\cdots\B(\vk_n,\t_n)\Big>'$},\nonumber\\
\label{deltadelta}\nonumber\\
\ee
where the primes indicate that one should remove the Dirac delta's coming from the momentum conservation.
If we now make the reasonable assumption that  the back-reaction of the magnetic fields on the DM evolution is negligible,  
we may write
$\delta_{{\rm c}L}(q,\t)=D(\t)/D(\t_{\rm in})\delta_{{\rm c}L}(q,\t_{\rm in})$, where $\tau_{\rm in}$ is some initial time and $D(\t)$ is the linear growth factor. If so, the consistency relation can be rewritten as

\be
\Big< \delta_{{\rm c}L}(\vq,\t)\B(\vk_1,\t_1)\cdots\B(\vk_n,\t_n)\Big>'_{q\to 0}
= -P_{{\rm c}L}(q,\tau)\sum_{a=1}^n \frac{D(\t_a)}{D(\tau)} \frac{{\vec q} \cdot \bk_a}{q^2}\Big<\B(\vk_1,\t_1)
\cdots\B(\vk_n,\t_n)\Big>',\nonumber\\
\label{deltadelta}\nonumber\\
\ee
where $P_{{\rm c}L}(q,\tau)$ is the linear DM power spectrum. 
For instance, for the three-point correlator, we obtain
\begin{eqnarray}
\Big< \delta_{{\rm c}L}(\vq,\t)\B(\vk_1,\t_1)\B(\vk_2,\t_2)\Big>'_{q\to 0}
&=&- P_{{\rm c}L}(q,\tau) \left(\frac{D(\t_1)}{D(\tau)}
-\frac{D(\t_2)}{D(\tau)}
\right)\nonumber\\
&\times& \frac{{\vec q} \cdot \bk_1}{q^2}\Big<\B(\vk_1,\t_1)\B(\vk_2,\t_2)\Big>'.
\label{deltadeltadelta1}
\end{eqnarray}
Let us check the validity of the consistency relation at second-order for the three-point correlator. From the induction equation (\ref{m3})
we deduce that at second-order  we have

\be
\B(\vx,\t)=\frac{\B_{\rm in}(\vx)}{a^2(\t)}+\frac{1}{a^2(\t)}\int^\t\,\d\eta\,a^2(\eta)\bnabla\times\left( \vec{v}_{{\rm c}L}(\vx,\eta)\times\frac{\B_{\rm in}(\vx)}{a^2(\eta)}\right),
\ee
where $\B_{\rm in}(\vx)$ is the initial condition for the magnetic field and we have considered it  a first-order perturbation. Also, we have replaced
in the induction equation $\vec{v}_{{\rm p}L}$ by $\vec{v}_{{\rm c}L}$. This amounts to ignoring  the magnetic back-reaction
and the relative motion of the baryons, effectively
considering a single fluid. We get

\be
B^i(\vk,\t)=\frac{1}{a^2(\t)}\left[B_{\rm in}^i(\vk)-\epsilon_{i\ell m}\,\epsilon_{mrs}\,
\int\frac{\d^3 q}{(2\pi)^3}\, \frac{1}{q^2} \left( q^\ell q^r + k^\ell q^r \right)\delta_{{\rm c}L}(\vq,\tau) B^s_{\rm in}(\vk-\vq)  \right].
\ee
Therefore at second-order we obtain

\begin{eqnarray}
a^2(\t_1)a^2(\t_2)\Big< \delta_{{\rm c}L}(\vq,\t)B^i(\vk_1,\t_1)B^j(\vk_2,\t_2)\Big>_{q\to 0}'&=&
\Big< \delta_{{\rm c}L}(\vq,\t)B^i_{\rm in}(\vk_1)B^j(\vk_2,\t_2)\Big>'_{q\to 0}\nonumber\\
&+&
\Big< \delta_{{\rm c}L}(\vq,\t)B^i(\vk_1,\t_1)B^j_{\rm in}(\vk_2)\Big>'_{q\to 0}\nonumber\\
&=&\epsilon_{j\ell m}\,\epsilon_{mrs}\,\frac{k_2^{\ell}q^{r}}{q^2}\Big<\delta_{{\rm c}L}(q,\tau)\delta_{{\rm c}L}(q,\tau_2)\Big>'\Big<B^i_{\rm in}(\vk_1)B^s_{\rm in}(\vk_2)\Big>'\nonumber\\
&+&\epsilon_{i\ell m}\,\epsilon_{mrs}\,\frac{k_1^{\ell}q^{r}}{q^2}\Big<\delta_{{\rm c}L}(q,\tau)\delta_{{\rm c}L}(q,\tau_1)\Big>'\Big<B^s_{\rm in}(\vk_1)B^j_{\rm in}(\vk_2)\Big>'.\nonumber\\
&&
\end{eqnarray}
Using the fact that 
\be
\epsilon_{i\ell m}\,\epsilon_{mrs}=-\epsilon_{m\ell i}\,\epsilon_{mrs}=-(\delta_{\ell r}\delta_{is}-
\delta_{\ell s}\delta_{ir})
\ee
 and the condition $k_iB_i^{\rm in}(\vk)=0$, we finally obtain

\begin{eqnarray}
a^2(\t_1)a^2(\t_2)\Big< \delta_{{\rm c}L}(\vq,\t)B^i(\vk_1,\t_1)B^j(\vk_2,\t_2)\Big>_{q\to 0}'
&=&-\frac{\vk_2\cdot\vq}{q^2}\Big<\delta_{{\rm c}L}(q,\tau)\delta_{{\rm c}L}(q,\tau_2)\Big>'\Big<B^i_{\rm in}(\vk_1)\B^j_{\rm in}(\vk_2)\Big>'\nonumber\\
&-&\frac{\vk_1\cdot\vq}{q^2}\Big<\delta_{{\rm c}L}(q,\tau)\delta_{{\rm c}L}(q,\tau_1)\Big>'\Big<B^i_{\rm in}(\vk_1)\B^j_{\rm in}(\vk_2)\Big>',\nonumber\\
&&
\end{eqnarray}
which reproduces the consistency relation in the soft limit for the three-point correlator.
Notice that, if  the correlators are computed all at equal times, the right-hand side of Eq. (\ref{deltadelta}) vanishes by momentum conservation and the $1/q^2$ infrared divergence will not appear when calculating invariant quantities. 
The vanishing of the equal-time correlators in the soft limit is  rooted in the fact  that one can locally eliminate the zero mode and the first spatial gradient of the  long and linear wavelength mode of the gravitational potential and that the response of the system on short scales is a uniform displacement.
Once more, we stress that these relations are valid  beyond linear order for the short wavelength modes 
which might well be in the non-perturbative regime. Furthermore, one could derive different consistency relations involving, for example, the correlations between the magnetic field and the plasma density contrast. 

\section{Violation of the cosmological magnetohydrodynamics consistency relations}
\noindent
The consistency relations we have found in the previous section are  due to 
``projection effects" as they are  a consequence of the change of coordinates induced by the long velocity mode.
This implies that one expects a violation of the consistency relation in the case in which the effect of the long DM mode on the initial conditions of the magnetic field is
not a coordinate transformation. 
If cosmic magnetic fields are  produced during inflation \cite{wreview}, this is precisely what happens. This is because they are likely to be correlated with the scalar curvature perturbation $\zeta$ which is  responsible for both the cosmic microwave background anisotropies and the large-scale structure.  Such  cross-correlation might   reveal the primordial nature of cosmic magnetic fields.  Indeed, consider a simple model where the interaction Lagrangian between the scalar field $\phi$ driving inflation and the electromagnetic field is of the form\footnote{The simplest gauge invariant models of inflationary magnetogenesis are known
to suffer from the problems of either large backreaction or strong coupling \cite{p1,p2}. For a possible solution see Ref. \cite{sol}.} 
\be
{\cal L}\supset f(\phi) F_{\mu\nu}^2\supset- \frac{f'}{{\cal H}}\zeta_L\,F_{\mu\nu}^2,
\ee
where we have used the fact that the comoving curvature perturbation is related to the linear 
inflation fluctuation by $\delta\phi_L=-(\phi'/{\cal H})\zeta_L$ (in the spatially flat gauge) and primes indicate differentiation with respect to the conformal time. Let us write the initial non-Gaussian initial condition for the magnetic field  as
\be 
\label{in}
\B_{\rm in}(\vx)=\B_{\rm g}(\vx)+ \frac{1}{2}b_{\rm NL}\zeta_L \B_{\rm g}(\vx),
\ee
where $\B_{\rm g}$ is the Gaussian initial condition for the magnetic field.  
One therefore gets $b_{\rm NL}={\cal O} (f'/f{\cal H})$ \cite{sloth1,sloth2}. In the squeezed limit, with the parametrization of the coupling of the form $f(\phi(\tau))\sim \tau^n$, the index $n$  can be related to the spectral index of the magnetic field power
spectrum $n_B=(4-2n$),  implying  $b_{\rm NL}=(n_B-4)$. In the
most interesting case of a scale-invariant magnetic  field spectrum $n_B = 0$, the non-linear
parameter is non-vanishing and given by $b_{\rm NL}=-4$. 

In the presence of such cross-correlations between the scalar perturbation $\zeta$ and the initial condition of the magnetic field,   we therefore expect  modulation effects from the curvature perturbation  long mode in 
the magnetic field power spectrum  and a corresponding violation of the consistency relation
at equal-time. This effect is not  easy to  gauge as  the short modes of the magnetic field are typically in the non-linear regime. However, if 
the primordial tiny seeds of the magnetic fields are to be amplified, one may consider a parametrization of the
magnetic field of the form
\be
\B\sim \frac{\B_{\rm in}}{a^2}\,{\rm exp}\left(\int^\tau\d\eta \,\gamma(\eta)\right),
\ee 
where $\gamma$ is a kind of (possibly directional) growth rate. 
This is exactly  what one finds when analyzing the  effects of mildly non-linear clustering
on magnetic fields frozen into a collapsing proto-cloud that
is falling into a CDM potential well \cite{bruni}. By ignoring  
the magnetic back-reaction on the baryons as well as
the relative velocity between baryons and DM, the authors of Ref. \cite{bruni} found that DM 
dominates the collapse and the gravitational anisotropy that
amplifies the magnetic field. The magnetic field loses one of its components and is confined
in the plane of the pancake and this  picture seems to be consistent
 with magnetic field observations in numerous spiral
and disk galaxies. At any rate, the important point is that the exponential amplification depends on the initial condition (\ref{in})  linearly, so that
\be
\frac{\delta \B}{\delta\zeta_L}\simeq \frac{1}{2}b_{\rm NL}\, \B.
\label{zz}
\ee
Maybe, a more robust argument is the following. Consider the induction equation
(\ref{m3}). Its formal solution is

\begin{eqnarray}
\B(\vx,\t)&=&\frac{\B_{\rm in}(\vx)}{a^2(\t)}+\frac{1}{a^2(\t)}\int^\t\,\d\eta\,a^2(\eta)\bnabla\times\left( \vec{v}_{{\rm p}}(\vx,\eta)\times\B(\vx,\eta)\right)\nonumber\\
&=&\frac{\B_{\rm in}(\vx)}{a^2(\t)}+\frac{1}{a^2(\t)}\int^\t\,\d\eta\,a^2(\eta)\bnabla\times\left( \vec{v}_{{\rm p}}(\vx,\eta)\times\frac{\B_{\rm in}(\vx)}{a^2(\eta)}\right)\nonumber\\
&+&\frac{1}{a^2(\t)}\int^\t\,\d\eta\,a^2(\eta)\bnabla\times\left[ \vec{v}_{{\rm p}}(\vx,\eta)\times\frac{1}{a^4(\eta)}\int^\eta\,\d\eta'\,a^2(\eta')\bnabla\times\left( \vec{v}_{{\rm p}}(\vx,\eta')\times\frac{\B_{\rm in}(\vx)}{a^2(\eta')}\right)
\right]\nonumber\\
&+&\cdots.
\end{eqnarray}
Working at linear order in the non-Gaussian parameter $b_{\rm NL}$, one can pick up  the cross-correlation with the curvature perturbation $\zeta_L$ either from the $\B_{\rm in}$ which explicitly appears in every term of the previous expression or from the one which appears implicitly in $\vec{v}_{{\rm p}}$.
In the first case, using Eq. (\ref{in}), one immediately recovers (\ref{zz}). The second contribution is more difficult to disentangle, but we expect it to be smaller as $\vec{v}_{{\rm p}}$ depends on the magnetic field quadratically. 

One has also to keep in mind that the dependence on  the  initial condition $\B_{\rm in}(\vx)$,  for the magnetic field at recombination, 
 is washed out  for comoving momenta larger than $k_{\rm d}$ because of    the dissipation of magnetic energy due to
the generation of MHD waves \cite{alf1,alf2,ruth}. Alfv\'en waves are the most effective in dissipating magnetic
energy, and damping occurs at scales $k^{-1}\simlt k_{\rm d}^{-1}=v_{\rm A} \tau$, 
 where $v_{\rm A}$ is the Alfv\'en speed.
Strictly speaking, Alfv\'en waves
are oscillatory perturbations superimposed on a homogeneous magnetic component  and the 
Alfv\'en speed depends
on the amplitude of the homogeneous component. In the cosmological context where the magnetic field is purely
stochastic, the amplitude of this component can be taken as the one of a low frequency component obtained by
smoothing the magnetic field amplitude over the scale $\sim k^{-1}$

\be
v_{\rm A}^2(k)=\frac{1}{2}\frac{\langle B^2\rangle_{k^{-1}}}{4\pi\left(\overline{P}_{\rm p}+\overline{\rho}_{\rm p}\right)},
\ee
where $\overline{\rho}_{\rm p}$ and $\overline{P}_{\rm p}$  indicate the background  energy density and pressure density of the plasma.  During the radiation epoch the 
Alfv\'en speed is a constant and the damping scale $k^{-1}_{\rm d}$ scales like the conformal time. During the matter-dominated epoch, the Alfv\'en speed decays like $a^{-1/2}\sim 1/\t$ and therefore the damping scale stops  growing. Taking a scale-invariant power spectrum for the
magnetic field, we estimate that at recombination the comoving  wavenumber is of the order of

\be
k_{\rm d}\simeq \frac{1}{v_{\rm A}(k)\t_{\rm rec}}\simeq \frac{3\cdot 10^2}{\t_{\rm rec}\,\ln k\t_{\rm rec}}\left(\frac{{\rm nG}}{B_0}\right)\simeq 70\left(\frac{{\rm nG}}{B_0}\right)\,{\rm Mpc}^{-1},
\ee
where $B_0$ is the present-day value of the large-scale magnetic field and $\t_{\rm rec}$ is the conformal time at recombination. 
 On scales larger than $k_{\rm d}^{-1}$ the magnetic field is not damped and it retains the information about its initial conditions when the modes of the magnetic field re-enter the Hubble radius till the recombination epoch.
  
The three-point correlator (\ref{deltadeltadelta1}) gets then modified to
\begin{eqnarray}
\Big< \delta_{{\rm c}L}(\vq,\t)\B(\vk_1,\t_1)\B(\vk_2,\t_2)\Big>'_{q\to 0}
&=&- P_{{\rm c}L}(q,\tau) \left(\frac{D(\t_1)}{D(\tau)}
-\frac{D(\t_2)}{D(\tau)}
\right) \frac{{\vec q} \cdot \bk_1}{q^2}\Big<\B(\vk_1,\t_1)\B(\vk_2,\t_2)\Big>'\nonumber\\
&+&\frac{5}{2}b_{\rm NL}\frac{{\cal H}(\t_0)\Omega(\t_0)}{q^2 T(q)}\frac{D(\tau_0)}{D(\tau)}\,P_{{\rm c}L}(q,\tau)\Big<\B(\vk_1,\t_1)\B(\vk_2,\t_2)\Big>',
\label{deltadeltadelta12}
\end{eqnarray}
where $T(q)$ is the DM linear transfer function and we have assumed that the effect of baryons are negligible,  so that
we can take $\zeta_L(q)=-(5/3)\Phi_L(q)=(5/3)\cdot(3/2)\,q^{-2}\,{\cal H}^2\Omega_{\rm c}\delta_{{\rm c}L}(q)$.
As expected, the equal-time three-point correlator is reduced to
\be
\label{aa}
\fbox{$\displaystyle
\Big< \delta_{{\rm c}L}(\vq,\t)\B(\vk_1,\t)\B(\vk_2,\t)\Big>'_{q\to 0}=\frac{5}{2}b_{\rm NL}\frac{{\cal H}(\t_0)\Omega(\t_0)}{q^2 T(q)}\frac{D(\tau_0)}{D(\tau)}\,P_{{\rm c}L}(q,\tau) \Big<\B(\vk_1,\t)\B(\vk_1,\t)\Big>'$}.
\ee
A measurement of a non-vanishing   equal-time  three-point correlator between the magnetic field and the large-scale DM fluctuation in the soft-limit   will be an indication
of the inflationary origin of the magnetic field.

We reiterate that, even though the exact form of the violation might be different from the one given in Eq. (\ref{aa}), any violation of the consistency relation at equal-time would be a sign of the fact that the primordial seeds for the magnetic fields were generated
during inflation. This is because no other origin can correlate with the long-wavelength DM fluctuations. 
As we mentioned in the introduction, there are two broad
classes of models for the origin of the seed fields: either the  seed fields are produced in the early universe, during epochs
preceding the structure formation, or  the seeds are created  during  the gravitational collapse leading to
structure formation. The second possibility would not explain a correlation with the DM fluctuations at very large scales, that is in the initial conditions.
The first possibility would also exclude causal mechanisms, such as the  one where the magnetic fields are generated during phase transitions:   their coherent lengths
will be much smaller than the horizon scales at a given epoch. The only option left which creates a correlation between the initial magnetic field and DM seeds is a common inflationary origin. 
All this  is worth further investigation.

\section{Conclusions}
In this paper we have focused on the symmetries enjoyed by the cosmological MHD equations. While we have certainly
not described all the possible symmetries, we have identified a subset of them which allow to write well-defined
consistency relations involving the soft-limit of $(n+1)$-correlators between the magnetic fields and the
DM fluctuations. These consistency relations have important properties: they vanish when the equal-time limit is considered, unless a cross-correlation between the magnetic field and the DM fluctuations already exists
on very large-scales, that is in the initial conditions. If a violation of the equal-time consistency relation is observed, this  would suggest an inflationary origin of the magnetic field seeds.

 \section*{Acknowledgments}
We thank R. Durrer, J.~Nore\~{n}a, and M. Sloth for very enlightening discussions and for reading the draft of the paper. The  research of A.K. was implemented under the ``Aristeia" Action of the 
``Operational Programme Education and Lifelong Learning''
and is co-funded by the European 
Social Fund (ESF) and National Resources.  A.K. is also   partially
supported by European Union's Seventh Framework Programme (FP7/2007-2013) under REA
grant agreement n. 329083. A.R. is supported by the Swiss National
Science Foundation (SNSF), project `The non-Gaussian Universe" (project number: 200021140236).

 \appendix
\section{Symmetries of the cosmological MHD in phase space}
\noindent
The Vlasov equation in physical coordinates $(\vec{r},t)$  is written as
 \begin{eqnarray}
 \frac{\del f}{\del t}+\frac{\del \vec{r}}{\del t}\cdot\bnabla_{\vec{r}}f+
 \frac{\del \vec{p}_r}{\del t}\bnabla_{\vec{p}_r}f=0,
 \end{eqnarray}
where $\vec{p}_r=m\dot{\vec{r}}$ and $f=f(\vec{p}_r,\vec{r},t)$ is the phase space density. 
In comoving coordinates $\vec{x}=\vec{r}/a$, with peculiar velocity $\vec{v}=a\dot{\vec{x}}$, and 
comoving momenta $p=ma^2\dot{\vec{x}}$ 
 we have the relations
 \begin{align}
 \frac{\del f}{\del t}\Big{|}_{\vec{p},\vec{x}}&=
 \frac{\del f}{\del t}\Big{|}_{\vec{p}_r,\vec{r}}+\left(m \ddot{a}\vec{x}-\frac{\dot{a}}{a^2} \vec{p}\right)\cdot\bnabla_{\vec{p}}f\Big{|}_{\vec{x},t}+
 \frac{\dot{a}}{a}\vec{x}\left(\bnabla_{\vec{x}}f\Big{|}_{\vec{p},t}-m\dot{a}\bnabla_{\vec{p}}f\Big{|}_{\vec{x},t}\right),
  \\
 \bnabla_{\vec{x}}\Big{|}_{\vec{p},t}f&=a
  \bnabla_{\vec{r}}\Big{|}_{\vec{p}_r,t}+m \dot{a}\bnabla_{\vec{p}}f\Big{|}_{\vec{x},t},\\
  \bnabla_{\vec{p}}f\Big{|}_{\vec{r},t}&=\bnabla_{\vec{p}_r}f\Big{|}_{\vec{r},t}.
\end{align}  
It can then be verified that the Vlasov equation in comoving coordinates is written as 
\begin{eqnarray}
 \frac{\del f}{\del t}+\frac{\vec{p}}{ma^2}\bnabla_{\vec{x}}f+
 \frac{\del\vec{p}}{\del t}\bnabla_{\vec{p}}f=0. \label{vlasov0}
 \end{eqnarray}
 \subsection{Vlasov-Poisson equations}
 In particular, in the case in which there is a gravitational force, the comoving momentum satisfies
 \begin{eqnarray}
 \frac{\d\vec{p}}{\d \t}=-ma\bnabla_{\vec{x}}\Phi(\vx,\t),
 \end{eqnarray}
where we have switched now to conformal time. Recall that $\Phi(\vx,\t)$ satisfies the Poisson equation
\be
\nabla^2\Phi(\bx,\t)=4\pi G\,  \overline \rho a^2
\delta(\bx,\tau) \label{Poisson}
 \ee
and therefore the Vlasov-Poisson equation is written as
\be\frac{\del}{\del \t}f(\vec{p},\bx,\t)
+\frac{1}{am}\vec{p}\cdot\bnabla_{\vec{x}}  f(\vec{p},\bx,\t)-am\bnabla_{\vec{x}}\Phi\cdot\bnabla_{\vec{p}}f(\vec{p},\bx,\t)=0. \label{vlasov1}
\ee
We will try to find symmetries of the Vlasov-Poisson equations (\ref{Poisson}) and (\ref{vlasov1}) of the form
\be
\t\to \t'=T(\t), ~~~\bx\to \bx'=\vy(\bx,\t), ~~~\vec{p}\to \vec{p}'=\vec{\Pi}(\vec{p},\bx,\t).
\ee
Since 
\begin{eqnarray}
&&\frac{\del}{\del \t}=\dT\frac{\del}{\del \t'}+\dot{y}^i\frac{\del }{\del y^i} +\dot{\Pi}^i\frac{\del }{\del \Pi^i},\\
&& \frac{\del }{\del x^i}=A_{ij}\frac{\del }{\del y^j}+B_{ij}\frac{\del }{\del \Pi^j},\\
&& \frac{\del }{\del p^i}=C_{ij}\frac{\del }{\del \Pi^j},
\end{eqnarray}
where 
\begin{eqnarray}
A_{ij}=A_{ij}(\bx,\t)=\frac{\del y^j}{\del x^i}, ~~~B_{ij}=B_{ij}(\bx,\t)=\frac{\del \Pi^j}{\del x^i}, ~~~C_{ij}=C_{ij}(\bx,\t)=\frac{\del \Pi^j}{\del p^i},\label{deriv}
\end{eqnarray}
we get from 
\be
 &&\frac{\del}{\del \t}f(\vec{p},\bx,\t)
+\frac{1}{a'm}p^i\frac{\del}{\del x^i}  f(\vec{p},\bx,\t)-a'(\t)m\frac{\del }{\del x^i}\Phi'\, \frac{\del}{\del p^i}f(\vec{p},\bx,\t)=0
 \label{vlasov3}
 \ee
the relation 
\be
 &&\dT\frac{\del}{\del \t'}f(\vec{p}',\bx',\t')+
\dot{y}^i\frac{\del}{\del y^i} f(\vec{p}',\bx',\t')+\dot{\Pi}^i\frac{\del }{\del \Pi^i}f(\vec{p}',\bx',\t')+\frac{1}{a' m}p^iA_{ij}\frac{\del}{\del y^j}f(\vec{p}',\bx',\t')\nonumber \\ &&
+\frac{1}{a'm} p^iB_{ij}\frac{\del }{\del \Pi^j}f(\vec{p}',\bx',\t')
 -a'(\t)m\frac{\del }{\del x^i}\Phi'C_{ij}\frac{\del }{\del \Pi^j}f(\vec{p}',\bx',\t')=0.
 \label{vlasov3}
\ee
The above equation gives the condition (vanishing of the $\del f/\del y^i$ term)
\begin{eqnarray}
\dT \Pi^i=\frac{a(\t')}{a'(\t)}A_{ij}p^j+a(\t')m\dot{y}^i, \label{pii}
\end{eqnarray}
whereas, from the vanishing of the $\del f/\del \Pi$ term, we get
\begin{eqnarray}
\frac{\del }{\del x^i}\Phi'(\bx,\t) =\frac{\dT a(\t')}{a'(\t)}C^{-1}_{ji}\frac{\del }{\del y^j}\Phi(\t',\bx')+\frac{1}{a'(\t)m}C^{-1}_{ij}
\left(\dot{\Pi}^j+
\frac{1}{a'(\t)m}p^k B_{kj}\right). \label{mod}
\end{eqnarray}
Now, we have the two following additional conditions to satisfy 
\begin{eqnarray}
 \overline \rho '(\t) a'(\t)^3=\overline \rho (\t')a(\t')^3, ~~~~
 \overline \rho '(\t) \Big{(}1+\delta'(\bx,\t)\Big{)}=\overline \rho (\t')
 \Big{(}1+\delta(\bx',\t')\Big{)}. \label{ra}
 \end{eqnarray} 
 Since 
 \begin{eqnarray}
 \rho(\bx,\t)=\frac{1}{a^3(\t)}\int \d^3 p\,f(\vec{p},\bx,\t)
 \end{eqnarray}
and 
\begin{eqnarray}
\overline \rho(\t)=\frac{1}{a^3(\t)^3}\int \d^3 x\d^3 p\,f(\vec{p},\bx,\t),
\end{eqnarray}
we get that 
\begin{eqnarray}
\det {\cal C}=\frac{a(\t')^3}{a'(\t)^3},   ~~~\det {\cal A}=\frac{a'(\t)^3}{a(\t')^3}, \label{det}
\end{eqnarray}
where ${\cal C}$ and  ${\cal A}$ are the $3\time 3$ matrices $(C_{ij})~(A_{ij})$, respectively.
Therefore, we find that $\overline \rho$ transforms as
\begin{eqnarray}
\overline \rho (\t)\to \overline \rho '(\t)=\frac{1}{\det{\cal A}}\, \overline \rho (\bx,\t).
\end{eqnarray}
Acting on Eq. (\ref{mod}) with $\del/\del x^i$ and using Poisson equation we get 
\begin{align}
4\pi G \overline \rho'(\t)a'(\t)^2\delta'(\bx,\t)&=\frac{\dT a(\t')}{a'(\t)}
C^{-1}_{ij} A_{ik}\frac{\del^2}{\del y^k \del y^j}\Phi(\bx',\t')+
\frac{1}{a'(\t)m}C^{-1}_{ij}\frac{\del }{\del x^i}\left(\dot{\Pi}^j+\frac{1}{a'(\t)m}p^kB_{kj}\right).
\end{align}
The only way this equation determines the transformation property of the overdensity $\delta$ is if
\begin{eqnarray}
C^{-1}_{ij} A_{ik}=M(\t) \delta_{jk},
\end{eqnarray}
which from (\ref{det}) gives that 
\begin{eqnarray}
M(\t)=(\det{\cal A})^{2/3}.
\end{eqnarray}
In this case we get 
\begin{eqnarray}
\delta'(\bx,\t)=\dT M(\t)\delta(\bx',\t')+\frac{C^{-1}_{ij}}{4\pi G \overline \rho'(\t)a'(\t)^3m}\frac{\del}{\del x^i}\left(\dot{\Pi}^j+\frac{1}{a'(\t)m}.p^kB_{kj}\right).
\end{eqnarray}
Then, from the second condition of Eq. (\ref{ra}) we get that 
\begin{eqnarray}
1+\delta'(\bx,\t)=\det{\cal A}(1+\delta(\bx',\t')),
\end{eqnarray}
so that 
\begin{eqnarray}
\dT M(\t)=\det {\cal A}, ~~~\frac{C^{-1}_{ij}}{4\pi G \overline \rho'(\t)a'(\t)^3m}\frac{\del}{\del x^i}\left(\dot{\Pi}^j+\frac{1}{a'(\t)m}p^kB_{kj}\right)=\det{\cal A}-1.
\label{con}
\end{eqnarray}
Therefore we find
\begin{eqnarray}
\dT=(\det{\cal A})^{1/3},
\end{eqnarray}
so that 
\begin{eqnarray}
\delta'(\bx,\t)=\det {\cal A}\, \delta(\bx',\t')+\det {\cal A}-1.
\end{eqnarray}
 Note that from Eq. (\ref{deriv}) we get that
\begin{eqnarray}
y^i=A_{ij}(\t)x^j+n^i(\t)\, , ~~~~\Pi^i=C_{ij}(\t)p^j+B_{ij}(\t) x^j+s^i(\t),
\end{eqnarray}
since from the second condition of Eq. (\ref{con}), if $B_{ij}$ was a function of $x^i$, then (\ref{con}) will depend on the momenta $p^i$ as well. Using that
\be 
\begin{aligned}
\Pi^i&=(\det{\cal A})^{-2/3}A_{ij}p^j+(\det{\cal A})^{-1/3}a(\t')m \dot{y}^i \\
&= (\det{\cal A})^{-2/3}A_{ij}p^j+(\det{\cal A})^{-1/3}a(\t')m \frac{\del}{\del\t} \left( A_{ij}x^j+n^i(\t)\right)
\end{aligned}
\ee
we can identify
\begin{equation}
B_{ij} = (\det\mathcal{A})^{-2/3} a'(\t)m\dot A_{ij}, ~~~~~ s^i(\t)=(\det\mathcal{A})^{-2/3} a'(\t)m\dot n^i(\t),
\end{equation}
and find that the second condition of (\ref{con}) is written as
\begin{eqnarray}
\frac{(\det \mathcal{A})^{2/3}}{4\pi G \overline \rho'(\t)a'(\t)^3}A^{-1}_{ij}\frac{\del}{\del \t}
\left(\frac{a'(\t)}{(\det{\cal A})^{2/3}} \frac{\del }{\del \t}A_{ij}\right)=\det{\cal A}-1. \label{aij}
\end{eqnarray}
This can be expanded to obtain
\be
\frac{1}{4\pi G\bar\rho'(\t)a'(\t)^2}\left[ \ddot A_{ij} + \left( \frac{\dot a'(\t)}{a'(\t)}-\frac{2}{3}\frac{\frac{\del}{\del \t}(\det\cal{A})}{\det \cal{A}}\right)\dot A_{ij}\right]+(1-\det\mathcal{A})A_{ij}=0
\ee
Collecting our results we find that Vlasov-Poisson system is invariant under the transformations
\begingroup
\addtolength{\jot}{1em}
\begin{align}
&\t\to \t'=\int^\t (\det{\cal A})^{1/3}\d\eta ,~~
x^i\to {x^i} '=A_{ij}(\t)x^j+n^i(\t), \label{yfl11}\\
&a'(\t)=(\det{\cal A})^{1/3}\,a(\t'),\label{ya1}\\ 
& \overline \rho '(\t)=\frac{1}{ \det{\cal A}}\,\overline \rho(\t') ,\label{yr1}\\ 
&\delta'(\bx,\t)=\det{\cal A}\, \delta(\bx,\t)+\det{\cal A}-1,\label{yfl21}\\ 
&p^i\to {p^i} '=
(\det{\cal A})^{-2/3}A_{ij}p^j+(\det{\cal A})^{-2/3}a'(\t)m 
\Big{(}\dot{A}_{ij}(\t)x^j+\dot{n}^i(\t)\Big{)}, \label{yfl31}
\\&
\frac{\del}{\del x^i}\Phi'(\bx,\t)
=(\det{\cal A})^{2/3}A^{-1}_{ij}\frac{\del}{\del y^j}\Phi(\bx',\t')
+\frac{1}{a'(\t)m}A^{-1}_{ij}\left((\det{\cal A})^{2/3}\dot{p}^{j'} +\dot A_{kj}p^k\right). \label{yfl41}
\end{align}
\endgroup
The transformations (\ref{yfl11})-(\ref{yfl41}) are generated by the matrix 
$A_{ij}(\t)$ which satisfies Eq. (\ref{aij}). As $A_{ij}$ has necessarily determinant non-zero, it belongs to the group GL$(3,\Re)$ of $3\times 3$ non-singular matrices. Every such matrix can be decomposed as 
\begin{eqnarray}
A_{ij}=\frac{1}{3}\Tr A \, \delta_{ij}+A_{(ij)}^T+A_{[ij]},
\end{eqnarray}
where
\begin{eqnarray}
A_{(ij)}^T=\frac{1}{2}\left(A_{ij}+A_{ji}\right)-\frac{1}{3}
\Tr A \delta_{ij}
\end{eqnarray}
and 
\begin{eqnarray}
A_{[ij]}=\frac{1}{2}\left(A_{ij}-A_{ji}\right)
\end{eqnarray}
are the tracless symmetric $A_{(ij)}^T$ and antisymetric parts $A_{[ij]}$ of any matrix $A_{ij}\in$ GL$(3,\Re)$.
Clearly the trace part $A_{ij}$ is the one that has been employed in the previous section. It is also clear that its antisymmetric part   generates vorticity since in this case
\begin{eqnarray}
(\bnabla \vec{p}')_{ij}=a'(\t)m (\det{\cal A})^{-1/3}
\dot{A}_{[ij]}(\t)
\end{eqnarray}
is non-zero if the antisymmetric part of $A_{ij}$ is non-vanishing. 
Notice that when  $\det {\cal A}=1$ the matrix $A_{ij}$ belongs to the special linear group SL$(3,\Re)$ and the transformations 
(\ref{yfl11})-(\ref{yfl41}) read

\begingroup
\addtolength{\jot}{1em}
\begin{align}
&\t\to \t'=\t ,~~
x^i\to {x^i} '=A_{ij}(\t)x^j+n^i(\t), \label{yfl110}\\
&a'(\t)=a(\t'),\label{ya10}\\ 
& \overline \rho '(\t)=\overline \rho(\t') ,\label{yr10}\\ 
&\delta'(\bx,\t)=\delta(\bx,\t),\label{yfl210}\\ 
&p^i\to {p^i} '=
A_{ij}p^j+a(\t)m 
\Big{(}\dot{A}_{ij}(\t)x^j+\dot{n}^i(\t)\Big{)}, \label{yfl310}
\\&
\frac{\del}{\del x^i}\Phi'(\bx,\t)
=A^{-1}_{ij}\frac{\del}{\del y^j}\Phi(\bx',\t')
+\frac{1}{a(\t)m}A^{-1}_{ij}\left(\dot{p}^{i'} + \dot A_{kj}p^k\right). \label{yfl410}
\end{align}
\endgroup
For $A_{ij}=\delta_{ij}$ we get the transformation used in this paper. 
However, we can still generate vorticity with a matrix $A_{ij}$ with  non-zero antisymmetric part.

\subsection{Vlasov-Poisson-Maxwell}

Let us now consider a collisionless plasma in the presence of gravitational, electric, and magnetic fields, $\vec{E}$ and $\vec{B}$. Let us assume that the plasma is composed of two fluids, of particles of masses $m_i$ and charges $e_i,~(i=1,2)$, electrons and baryons, with corresponding particle distribution functions $f_i(\vec{p},\vec{x},t)$, and which obey the Vlasov-Poisson-Maxwell equation
\begin{eqnarray}\frac{\del}{\del \t}f_i(\vec{p},\bx,\t)
+\frac{\vec{p}_i}{am_i}\cdot\bnabla_{\vec{x}}  f_i(\vec{p},\bx,\t)-\Big{[}am_i\bnabla_{\vec{x}}\Phi- a e_i\left(\vec{E}+
\frac{\vec{p}_i}{a m_i}\times \vec{B}\right)
\Big{]}\cdot\bnabla_{\vec{p}}f_i(\vec{p},\vx,\t)=0. 
 \label{vlasov10}
\end{eqnarray}
Note that $\vec{B}=\vec{B}_{\vx}$ and $\vec{E}=\vec{E}_{\vx}$ are the electric and magnetic fields  in the  comoving frame and they are related to the corresponding fields $\vec{B}_{\vec{r}}$ and $\vec{E}_{\vec{r}}$ in the  coordinate frame by
 \begin{eqnarray}
 \vec{E}_{\vx}=a\vec{E}_{\vec{r}}+\dot{a}\, \vec{r}\times\vec{B}_{\vec{r}}, ~~~\vec{B}_{\bx}=a\vec{B}_{\vec{r}}.
 \end{eqnarray}
As usual, we define the number  density field of the fluids 
\begin{eqnarray}
n_i(\vec{x},t)= a^{-3}\int \d^3p f_i 
\end{eqnarray}
and the mean velocity of the fluids  
\begin{eqnarray}
\vec{v}_i(\vec{x},t)=\frac{\int \d^3p \frac{\vec{p}_i}{m_i a} f_i}{\int \d^3p f_i}.
\end{eqnarray}
By taking moments of the Vlasov-Poisson-Maxwell equation, we obtain the evolution equations for the corresponding moments. For example, the zeroth-order moment gives  
 \begin{eqnarray}
 \frac{\del (n_ia^3)}{\del \t}+\bnabla(a^3n_i \vec{v}_i)=0.
 \end{eqnarray}
Similarly, the first moment equation gives 
\begin{eqnarray}
\frac{\del (a^3\rho_i \vec{v}_i)}{\del \t}+a^3\left( \mathcal{H}\rho_i\vec{v}_i+\vec{v}_i\bnabla(\rho_i\vec{v}_i)+\rho_i(\vec{v}_i\cdot\bnabla)\vec{v}_i+\rho_i\vec{\nabla}\Phi-
e_in_i\left(\vec{E}+\vec{v}_i\times \vec{B}\right)+\bnabla P_i \right)=0,
\end{eqnarray}
for the case of homogeneous and isotropic collapse, where $P_i$ is the pressure and $
\rho_i=m_i n_i$.
Summing over the fluids  and defining the plasma mass density function $\rho_{\rm p}$, the velocity $\vec{v}_{\rm p}$ and the pressure as 
 \begin{eqnarray}
 \rho_{\rm p}=\frac{m_1 \rho_1+m_2 \rho_2}{m_1 + m_2} , ~~~\vec{v}_{\rm p}=\frac{\rho_1 \vec{v}_1 + \rho_2 \vec{v}_2  }{\rho_1 + \rho_2}, ~~~P_{\rm p}=P_1+P_2, 
 \end{eqnarray}
as well as the charge density $\rho_e$ and  the current density $\vec{j}_{\rm p}$ as  
\begin{eqnarray}
\rho_e=\sum_ie_in_i=e(n_1-n_2), ~~~\vec{j}_{\rm p}=\sum_i e_in_iv_i=e(n_1\vec{v}_1-n_2\vec{v}_2), 
\end{eqnarray}
we find the plasma continuity equation \cite{ma}
\begin{eqnarray}
 \frac{\del (\rho_{\rm p} a^3)}{\del \t}+\bnabla(a^3\rho_{\rm p}\vec{v}_{\rm p})=0,
 \end{eqnarray}
 as well as momentum conservation 
 \begin{eqnarray}
 \rho_{\rm p} \left( \frac{\del \vec{v}_{\rm p}}{\del \t}+\mathcal{H}\vec{v}_{\rm p}+(\vec{v}_{\rm p}\cdot\bnabla)\vec{v}_{\rm p}+\vec{\nabla}\Phi \right)-
\rho_e\vec{E}-\vec{j}_{\rm p}\times \vec{B}+\bnabla P_{\rm p}=0. \label{e-e}
 \end{eqnarray}
 In the last equation we have assumed $m_1\gg m_2$; this is a good approximation because ultimately 
 the two particles have to be identified with protons (ions) and electrons and therefore one can ignore the 
  contribution of the light electrons to the non-linear advection term. 
 In addition, in MHD we assume quasi-neutrality 
\be 
\rho_e=0 \label{r1}
\ee 
 and ideal Ohm's law, {\it i.e.}  zero resistivity,  so that 
 $\vec{E}=\vec{B}\times \vec{v}_{\rm p}$ in the physical frame. Then Maxwell's equation 
\begin{eqnarray}
 \bnabla\times\vec{B}-\frac{1}{c^2}\frac{\del \vec{E}}{\del t}=\mu \vec{J}_{\rm p}
 \end{eqnarray} 
 turns out to be 
 \begin{eqnarray}
 \bnabla\times\vec{B}+\frac{1}{c^2}\frac{\del( \vec{v}_{\rm p} \times \vec{B})}{\del t}=\mu \vec{J}_{\rm p}.
 \end{eqnarray}
 Therefore, in the non-relativistic limit $v\ll c$, the displacement current vanishes and we are left with Ampere's law which is written as 
 \be 
 \mu\vec{j}_{\rm p}=\frac{1}{a^2}\bnabla\times \vec{B},\label{ampere}
 \ee
in the comoving frame. By plugging  Eqs. (\ref{r1}) and  (\ref{ampere})  in to Eq.  (\ref{e-e}), and returning $\vec{B}$ to its physical value, we get the plasma momentum conservation (\ref{m2}).
  
It is now straightforward to verify that the Vlasov-Poisson-Maxwell  equation (\ref{vlasov10}) is invariant under the transformations (\ref{yfl110})-(\ref{yfl410}). Indeed, 
we have shown already that this system is invariant for $\vec{B}=\vec{E}=0$. Thus (\ref{vlasov10}) is invariant for non vanishing $\vec{E}$ and $\vec{B}$ if 
the term 
\be
I_i=ae_i(\vec{E}+\vec{v}_i\times \vec{B})
\cdot\bnabla_{\vec{p}}f(\vec{p},\bx,\t)
\label{ext0000}
\ee
is also invariant. In other words, we must have

\begin{align}
a(\t')e_i\Big{(}\vec{E}(\vec{x}',\t')+\frac{\vec{p}_i'}{ma(\t')}\times \vec{B}(\vec{x}',\t')\Big{)}
\cdot\bnabla_{\vec{p}'}f(\vec{p}',\bx',\t')=
a(\t)e_i\Big{(}\vec{E}'(\vec{x},\t)+\frac{\vec{p}_i}{ma'(\t)}\times \vec{B}'(\vec{x},\t)\Big{)}
\cdot\bnabla_{\vec{p}}f(\vec{p},\bx,\t)\nonumber\\
\end{align}
in order for the Vlasov-Poisson-Maxwell equation (\ref{vlasov10}) to be invariant.
Then, by using (\ref{yfl11})-(\ref{yfl41}) it is straightforward to verify that (\ref{ext0000}) specifies the transformation properties of the electric and magnetic fields to be
\begin{eqnarray}
&&E'_i(\vx,\t)=(\det{\cal A})^{1/3} A_{ij}^{-1} E^j(\vx',\t')
+\epsilon_{jk\ell}\big{(}\dot{A}_{kn}x^n+\dot{n}^k\big{)}B^{\ell}(\vx',\t')A_{ij}^{-1},\label{elec}\\
&&B'_i(\vx,\t)=\frac{1}{2}\epsilon_{ijk}\epsilon_{q\ell m}A_{jq}^{-1}A_{\ell k} B^m(\vx',\t').\label{magn}
\end{eqnarray}
For the consistency relation described in the text  we have employed the  particular case of the 
general transformations (\ref{yfl11})-(\ref{yfl41}), (\ref{elec}) and (\ref{magn}), namely 
\begin{eqnarray}
T(\t)=\t, ~~~A_{ij}=\delta_{ij}.
\end{eqnarray} 
 In this case, (\ref{elec}) and (\ref{magn}) turn out to be
 
 \begin{eqnarray}
&& \vec{E}'(\vx,\t)=\vec{E}(\vx',\t')+\dot{\vec{n}}\times \vec{B}(\vx',\t'),\\ 
&&\vec{B}'(\vx,\t)=\vec{B}(\vx',\t').
 \end{eqnarray}
 which is the transformation we employed in section 2. 
 Of course, since the electric field is multiplied by the charge density  $\rho_e$  and the latter vanishes for a quasi-neutral plasma, it does not enter in the momentum conservation equation.


\end{document}